\documentclass[sigconf]{acmart}

\usepackage{hyperref}
\usepackage{url}
\usepackage{booktabs}
\usepackage{array,multirow,graphicx}
\usepackage{tabularx,booktabs}
\newcolumntype{Y}{>{\centering\arraybackslash}X}
\usepackage{listings}

\usepackage{amsmath,amsfonts,bm}

\def\eqref#1{equation~\ref{#1}}

\def\1{\bm{1}}

\def\vc{{\bm{c}}}

\def\vk{{\bm{k}}}

\def\vq{{\bm{q}}}

\def\vz{{\bm{z}}}

\DeclareMathAlphabet{\mathsfit}{\encodingdefault}{\sfdefault}{m}{sl}
\SetMathAlphabet{\mathsfit}{bold}{\encodingdefault}{\sfdefault}{bx}{n}

\newcommand{\sigmoid}{\sigma}

\newcommand{\gorshochek}{\texttt{CodeTransformator}\xspace}

\usepackage{xspace}
\newcommand{\ie}{\textit{i.e.,}\xspace}
\newcommand{\eg}{\textit{e.g.,}\xspace}

\usepackage{hyperref}
\usepackage{url}
\usepackage{bbm}
\usepackage{booktabs}
\usepackage{thmtools} 
\usepackage{thm-restate}
\usepackage{graphicx}
\usepackage{array,multirow}

\usepackage{graphicx}
\usepackage{caption}
\usepackage{subcaption}
\graphicspath{ {./figures/} }
\usepackage{thmtools} 
\usepackage{thm-restate}
\usepackage{tikz}
\usepackage{cleveref}
\usepackage{comment}

\AtBeginDocument{%
  \providecommand\BibTeX{{%
    \normalfont B\kern-0.5em{\scshape i\kern-0.25em b}\kern-0.8em\TeX}}}

\begin{document}

\title{Evaluation of Contrastive Learning with Various Code Representations for Code Clone Detection}

\author{Maksim Zubkov}
\email{maksim.zubkov@epfl.ch}
\affiliation{%
  \institution{Swiss Federal Institute of Technology (EPFL)}
  \country{}
}
\author{Egor Spirin}
\email{spirin.egor@gmail.com}
\affiliation{%
  \institution{JetBrains Research}
  \country{}
}
\author{Egor Bogomolov}
\email{egor.bogomolov@jetbrains.com}
\affiliation{%
  \institution{JetBrains Research}
  \country{}
}
\author{Timofey Bryksin}
\email{timofey.bryksin@jetbrains.com}
\affiliation{%
  \institution{JetBrains Research}
  \country{}
}



\renewcommand{\shortauthors}{Trovato and Tobin, et al.}

\begin{abstract}
Code clones are pairs of code snippets that implement similar functionality. Clone detection is a fundamental branch of automatic source code comprehension, having many applications in refactoring recommendation, plagiarism detection, and code summarization. A particularly interesting case of clone detection is the detection of semantic clones, \ie code snippets that have the same functionality but significantly differ in implementation. A promising approach to detecting semantic clones is contrastive learning (CL), a machine learning paradigm popular in computer vision but not yet commonly adopted for code processing.

Our work aims to evaluate the most popular CL algorithms combined with three source code representations on two tasks. The first task is code clone detection, which we evaluate on the \textsc{POJ-104} dataset containing implementations of 104 algorithms. The second task is plagiarism detection.  To evaluate the models on this task, we introduce \gorshochek, a tool for transforming source code. We use it to create a dataset that mimics plagiarised code based on competitive programming solutions. We trained nine models for both tasks and compared them with six existing approaches, including traditional tools and modern pre-trained neural models.
The results of our evaluation show that proposed models perform diversely in each task, however the performance of the graph-based models is generally above the others. Among CL algorithms, SimCLR and SwAV lead to better results, while Moco is the most robust approach. Our code and trained models are available at \url{https://doi.org/10.5281/zenodo.6360627}, \url{https://doi.org/10.5281/zenodo.5596345}.
\end{abstract}



\keywords{Source Code Comprehension,
Clone Detection,
Contrastive Learning,
Source Code Representation}

\maketitle

\section{Introduction}\label{sec:introduction}

An open problem in the software engineering (SE) domain is the building of systems that accurately detect code similarity. Such systems aim to determine whether code snippets solve a similar (or equivalent) problem, even if the snippets' implementation, language, program input, or output structure are different. Finding similar code fragments confidently turns out to be very useful in various SE tasks, such as refactoring recommendation~\cite{aniche2020effectiveness}, detection of duplicates~\cite{wang2020detecting}, bugs~\cite{9251083}, vulnerabilities~\cite{zhou2019devign}, and cross-language code search~\cite{bui2020infercode, zugner2021languageagnostic}. 

An important case of code similarity detection is the detection of code clones, \ie code snippets with identical functionality. Code clones are commonly divided into four types~\cite{svajlenko2020survey}, with \textsl{Type~IV} being the hardest to detect. \textsl{Type~I-III} clones refer to gradually increasing differences in the implementation (\eg changed names of tokens, order of operations, insertion of dead code). Meanwhile, code fragments that constitute \textsl{Type~IV} clones have completely different implementations, only sharing the functionality. Detection of such clones is hard as it requires algorithms to capture the code's internal semantics.


Detection of essentially identical objects arises in other domains: for example, in computer vision (CV) to find images of the same object or in natural language processing (NLP) to find texts with the same meaning. Both in these domains and in code clone detection, researchers commonly employ machine learning (ML) techniques to identify such objects. In particular, the application of contrastive learning (CL) approaches is of great interest as in this task it shows superior results compared to other ML approaches~\cite{chen2020mocov2, chen2020simple, chen2021empirical, giorgi2021declutr, rethmeier2021primer, xie2021selfsupervised}.

Previous works that studied applications of CL approaches for the code clone detection task focused only on one classical contrastive learning method each~\cite{jain2020contrastive,ye2021misim}. Given the recent progress in the contrastive learning methods in CV and NLP, there is a need to evaluate modern CL approaches on the clone detection task. Additionally, an important choice one should make when applying contrastive learning to code is how to represent code snippets before passing them to an ML model: as text~\cite{feng2020codebert, jain2020contrastive}, as an abstract syntax tree~\cite{bui2020infercode, alon2018codeseq}, as a more complex graph structure~\cite{zhou2019devign}, or use a custom representation~\cite{zuegner_code_transformer_2021}. Thus, an evaluation should not only compare CL approaches, but also combinations of contrastive learning methods and code representations.

With this work, we evaluate combinations of three widely adopted contrastive learning methods and three code representations on the code clone detection task. We use two datasets of C/C++ code: \textsc{POJ-104}~\cite{mou2015convolutional} that contains algorithms implemented by different programmers and a dataset of solutions to competitive programming problems. We also present a tool called \gorshochek which we use to augment the latter dataset by transforming code snippets into functionally equivalent ones. We compare the studied models with multiple baselines, both ML-based and traditional (non-ML) token-based ones. 

The results we obtain partially agree with the conclusions drawn in the research done by Han et al.~\cite{han2021comparison}. The evaluation results show that models utilizing graph representation of code are generally more accurate than models that treat code as raw text only. Another interesting finding is that Transformer-based models are much more sensitive to the choice of hyperparameters than other models we experiment with. Overall, we develop and make publicly available a unified, extensible framework for enabling easy side-by-side comparison of various encoders and contrastive learning algorithms on the code clone detection task.

The contributions of this work are:
\begin{itemize}
    \item Comparison of multiple contrastive learning approaches, namely Moco~\cite{chen2020mocov2}, SimCLR~\cite{chen2020simple}, and SwAV~\cite{caron2020unsupervised}, combined with three different code representations on two datasets for the code clone detection task.
    \item \gorshochek, an extensible open-source tool for augmenting datasets for the plagiarism detection task by applying functionality-preserving transformations.
    \item An open-source framework that enables comparison of CL approaches powered by different code representations.
\end{itemize}

\smallskip
The rest of this paper is organized as follows. \Cref{sec:CL} describes the concept of contrastive learning and presents CL algorithms and code representation models that we evaluate in this work. \Cref{sec:setup} describes the experimental setup: tasks, datasets, and metrics that we use. In \Cref{sec:implementation} we present implementation details of the framework we developed to compare the models, as well as model configurations that we use in each experiment. In \Cref{sec:results} we discuss the evaluation results. \Cref{sec:related-work} summarizes the related studies and compares them to our work. Finally, in \Cref{sec:conclusion} we conclude and outline possible directions for the future work.

\section{Contrastive Learning Approach}\label{sec:CL}

Contrastive learning (CL) is a machine learning paradigm used to train models to identify similar and distinguish dissimilar objects. CL has demonstrated impressive progress in self-supervised learning\footnote{A group of machine learning methods that are able to train on an unlabeled dataset.} for various domains, such as CV, NLP, graph representation learning, and audio processing~\cite{le2020contrastive}.
In this paradigm, the model is trained to produce embeddings (\ie numerical vectors) for objects such that similar objects will have close embeddings and different objects, respectively, will have distinctive ones.
Therefore, the idea of applying this approach to code clone detection task seems natural.
In the end, similar programs will have close embeddings, so the decision whether two programs are clones or not is determined by the distance between two vectors.

Driven by the rapid development of deep learning, the number of contrastive learning algorithms has grown substantially in recent years.
In this work, we focus on the three most successful and popular CL algorithms that have shown good results in recent works~\cite{le2020contrastive} and explore their core differences. 

\begin{figure}[!th]
\begin{center}
    \includegraphics[width=9.5cm]{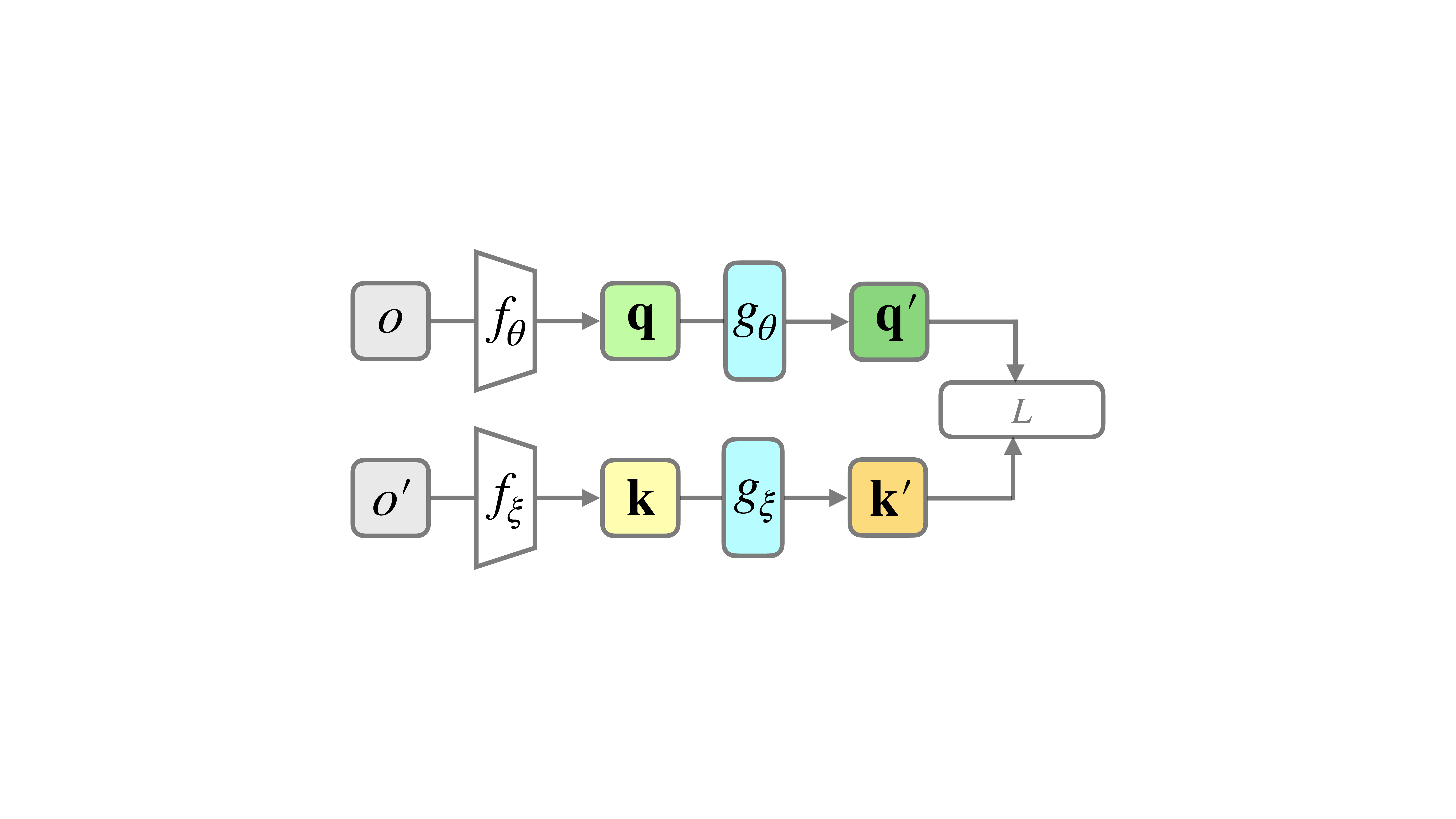}
\end{center}
\caption{Illustration of the general contrastive learning pipeline. $L$ is the objective function we aim to optimize during training.}
\label{fig:cl}
\end{figure}

\Cref{fig:cl} shows the common pipeline of a CL algorithm.
That is, given two different objects $o$ and $o'$, we pass them into an \textit{encoder} model $f$ and obtain their embeddings $\vq$ and $\vk$.
We further pass these embeddings into a special \textit{projector} model $g$ which is specific for each CL algorithm. In a simple case, $g$ can be an identical function, \ie just pass $\vq$ and $\vk$ onwards.
After the transformation by the projector, we receive vectors $\vq'$ and $\vk'$. Based on the transformed vectors, we compute the loss function (\eg  Noise Contrastive Estimation loss~\cite{Gutmann2010NoisecontrastiveEA}) that is optimized during the training stage.
As in other deep learning approaches, we update model parameters through the backpropagation algorithm to optimize the loss function.

Thus, for training, we need pairs of two objects: a target object $o$ and a reference object $o'$.
During the training process, we alternately choose $o'$ to be either a positive or a negative example, \ie to be equivalent or distinct to $o$, respectively.
The choice of both positive and negative examples is very important to train a robust model.
The more diverse and complex training pairs are, the more robust model we get after training.

In order to make a prediction with a trained model, we use the encoder model $f$ to build embeddings of target objects and then compute the distance between embeddings to determine whether the respective objects are clones or not. 
Thus, we use projector model $g$ only during the training phase in order to improve the performance and convergence of CL algorithms.

\subsection{Contrastive Learning Algorithms}

In this work, we focus on three CL algorithms that recently proved to be useful in the CV and NLP domains.
\smallskip

\subsubsection{Simple Framework for Contrastive Learning of Visual Representations (SimCLR)~\cite{chen2020simple}}

\begin{figure}[!th]
\begin{center}
    \includegraphics[width=9cm]{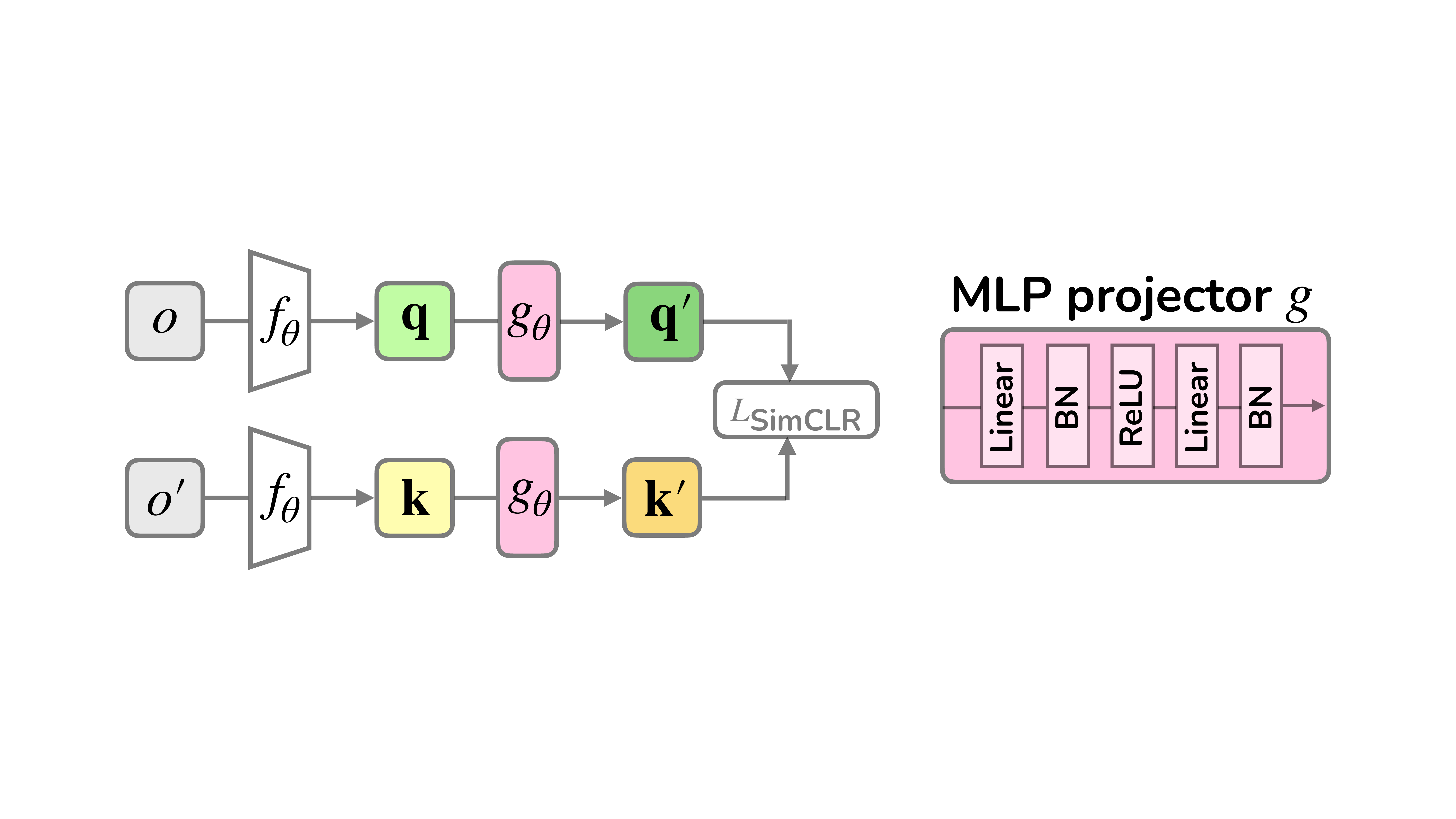}
\end{center}
\caption{Illustration of the SimCLR approach~\cite{chen2020simple}.}
\label{fig:simclr}
\end{figure}

SimCLR is a fairly simple yet effective CL approach that combines a number of ideas proposed in earlier works.
\Cref{fig:simclr} presents the overview of this algorithm.
A distinctive feature of SimCLR is the usage of a projector model $g$ represented by a multilayer perceptron.
The projector allows to separate the embeddings learned by the model to solve the optimization problem (minimizing the loss function) from the embedding generated by the encoder model.

To collect negative examples, SimCLR uses the following mechanism.
Given a batch of clone pairs $\{\langle o_i; p_i \rangle \}_{i=1}^B$, where $p_i$ is an object similar to $o_i$ (\ie a positive example), SimCLR considers other target objects $\{o_i\}_{i\not=j}$ to be negative examples for the given one $o_j$. This way, the algorithm makes an assumption that all the target objects $o_i$ in the batch are dissimilar.

\subsubsection{Momentum Contrast for Unsupervised Visual Representation Learning (Moco)~\cite{chen2020mocov2}}

\begin{figure}[!th]
\begin{center}
    \includegraphics[width=9cm]{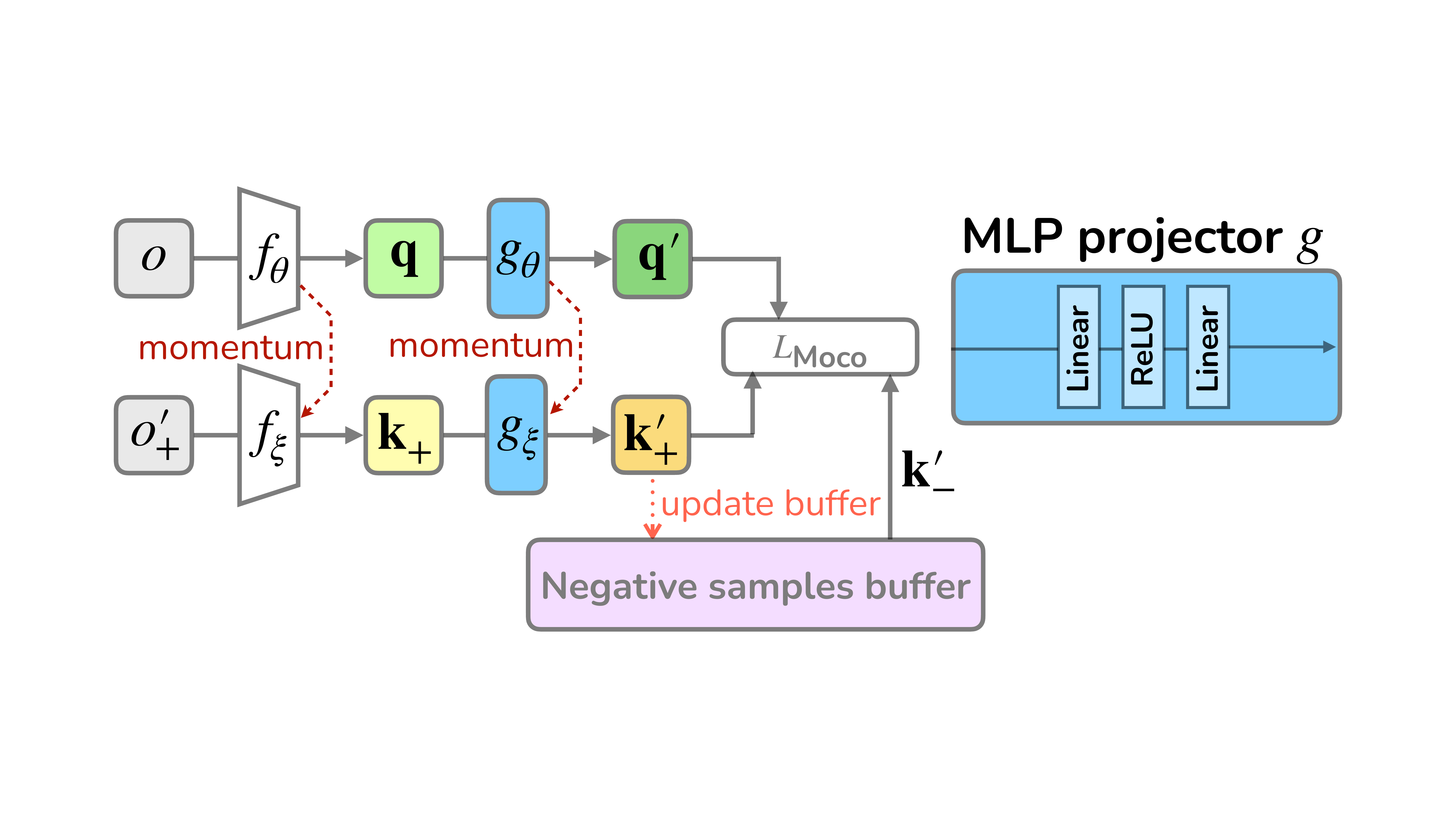}
\end{center}
\caption{Illustration of the Moco approach~\cite{chen2020mocov2}. }
\label{fig:moco}
\end{figure}

Moco is another popular approach to train encoder networks in the CL paradigm.
Similar to SimCLR, Moco was originally developed in the computer vision domain. However, it has already shown good results when applied to source code~\cite{jain2020contrastive}.
\Cref{fig:moco} shows the overview of the Moco approach.

One of the key differences between Moco and SimCLR is the buffer, which essentially is a queue of processed examples.
At each training step, there is a batch of object embeddings and their respective positive examples.
To collect negative examples, Moco uses embeddings of objects from previous steps that are stored in a queue.
At the end of each training step, Moco adds vector representations $\{\vq'_{i}\}_{i=1}^B$, where $B$ is the batch size, to the end of the queue.

The second distinguishing feature of Moco is the way it operates with embeddings.
This approach uses the same architectures $f$ and $g$ for building embeddings $\vk$ and $\vk'$ respectively, but at each step it updates their parameters using the momentum, which is essentially a moving average:
\begin{gather}
f_{\xi} = m \cdot f_{\xi} + (1-m) \cdot f_{\theta}, \\
g_{\xi} = m \cdot g_{\xi} + (1-m) \cdot g_{\theta},
\end{gather}
where $m \in [0,1)$ is a hyperparameter of the model.
Backpropagation is only used to update the models $f_{\theta}$ and $g_{\theta}$, which are the models for processing target objects. This way, $f_{\xi}$ and $g_{\xi}$ only approximate the parameters of the main models. Therefore, embeddings of reference objects are slightly different from the embedding of the target object, making the model more robust to changes in the reference objects.

\subsubsection{Swapping Assignments between multiple Views (SwAV)~\cite{caron2020unsupervised}}

\begin{figure}[!th]
\begin{center}
    \includegraphics[width=9cm]{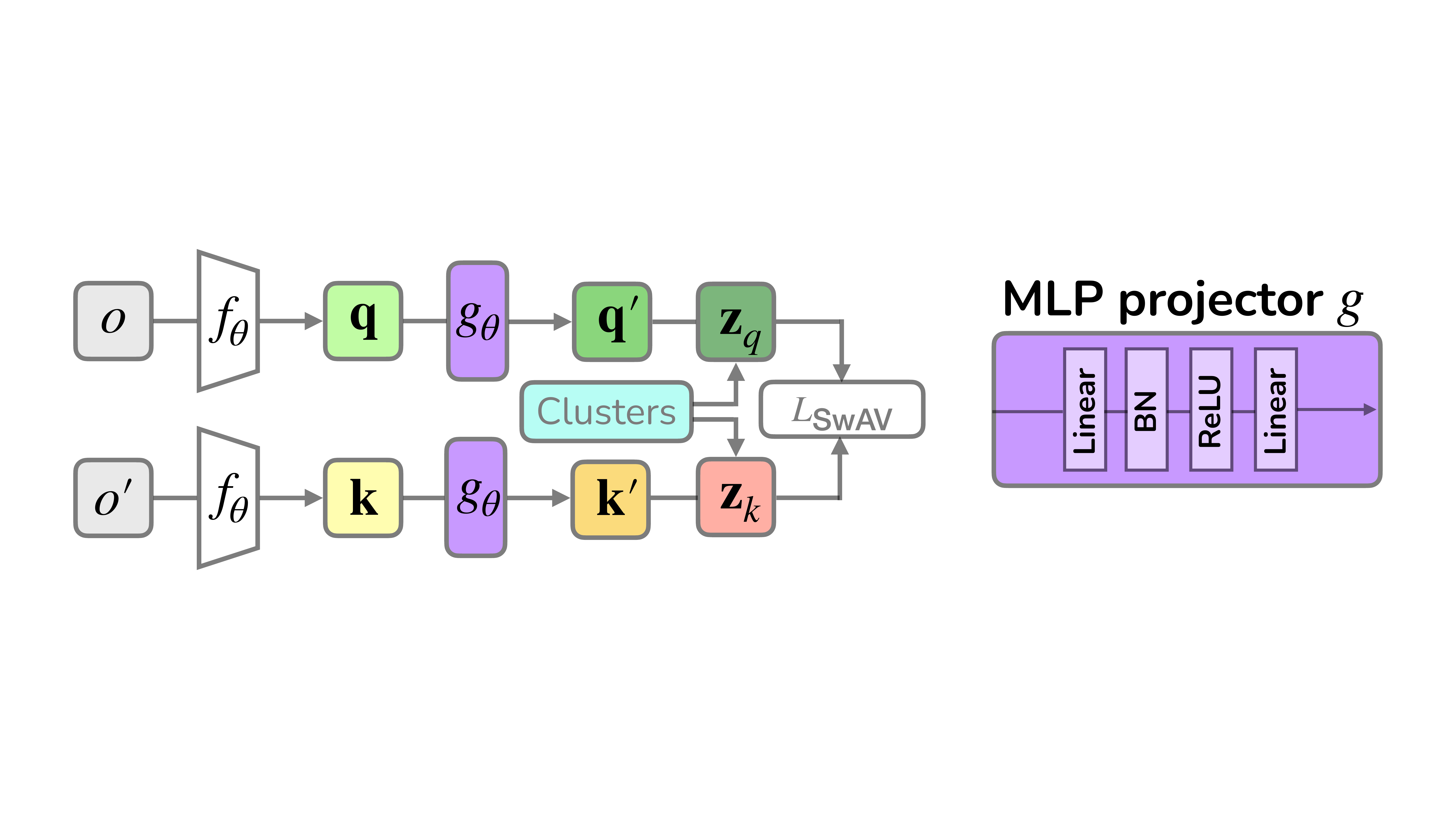}
\end{center}
\caption{Illustration of the SwAV approach~\cite{caron2020unsupervised}.}
\label{fig:swav}
\end{figure}

SwAV reformulates the representation extraction task as online clustering. \Cref{fig:swav} shows an overview of the SwAV approach.

The authors introduce a set of $L$ trainable vectors $\{\vc_1, \vc_2, ..., \vc_L\}$, called prototypes, that may be considered as the clusters in which the dataset should be partitioned.
After computing the embeddings $\vq'$ and $\vk'$, the algorithm decomposes them into weighted sums of $\{\vc_1, \vc_2, ..., \vc_L\}$.
The coefficients of these decompositions $\vz_{q}$ and $\vz_k$ are then used to calculate the loss function.

A major difference of SwAV compared to the previously described approaches is that it does not use negative examples.

\smallskip

All in all, there are many possible approaches to train object representations in the contrastive learning paradigm.
They differ in the way of collecting negative examples, processing intermediate embeddings, and choosing loss functions.
Existing works that applied CL in the SE did not compare different CL approaches and rather focused on a single algorithm each. Thus, it remains an open question which CL approaches are more suitable for SE tasks. In this work, we study the code clone detection task.

\subsection{Representation of Source Code in Neural Networks}

The first step of contrastive learning algorithms is the transformation of objects into numerical vectors, or embeddings. This transformation is usually done using a neural network.
For the code clone detection task, the objects are snippets of source code: functions, classes, or even complete files.

In order to transform source code into an embedding, we need to represent the code fragment in a way suitable for a neural network. These representations differ in the way they treat code: as plain text, as an abstract syntax tree (AST), or as a more complex graph structure. Depending on the representation, types of neural networks that we can use also vary. In the rest of this subsection we describe different kinds of code representations and models that we will use with them.

\newdimen\figrasterwd
\figrasterwd\textwidth

\begin{figure*}[ht]
    \centering
    \parbox{0.71\figrasterwd}{
        \parbox{.2\figrasterwd}{
            \subcaptionbox{
                An example of a code snippet. Text representation of code uses it directly after splitting into tokens.
                \label{fig:snippet}
            }{
                \includegraphics[width=\hsize]{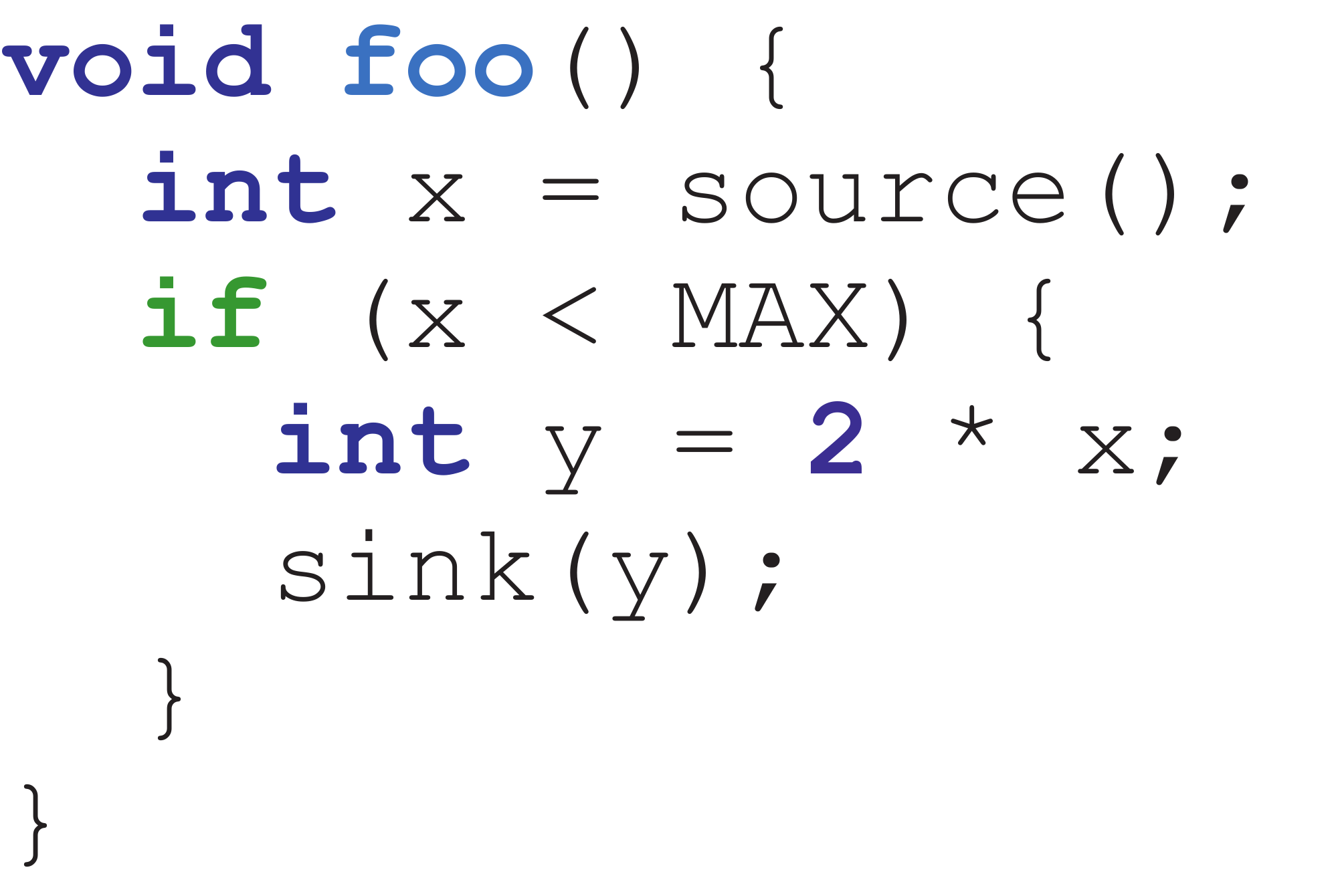}
            }
        }
        \hskip1em
        \parbox{.47\figrasterwd}{%
            \subcaptionbox{
                The corresponding graph representation with edges from AST, CFG, and PDG.
                \label{fig:graph}
            }{
                \includegraphics[width=\hsize]{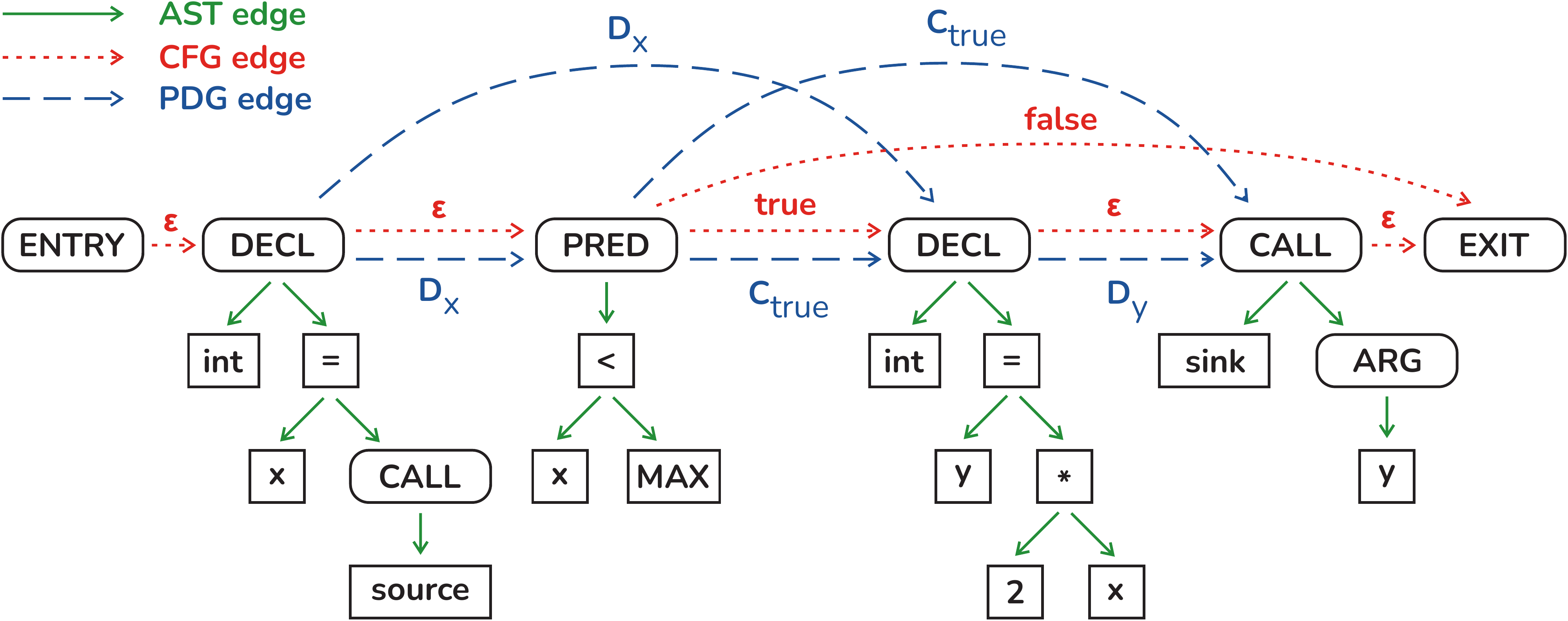}
            }
        }
    }
\caption{An example of a code snippet and more complex representations build from it.}
\end{figure*}

\subsubsection{Text-based Representation of Code, BERT}

Raw text is a natural representation of source code.
It was originally used in most of the early works on program analysis and still remains popular as it allows direct application of methods from the NLP domain while being rather easy to use~\cite{allamanis2015suggesting, theeten2019import2vec,feng2020codebert, jain2020contrastive}.
\Cref{fig:snippet} shows an example of a code snippet. In text representation, we split the snippet into tokens, and treat tokens as if they were words in a text written in natural language. 

We use BERT~\cite{devlin2019bert} as an encoder model that works with text representation of code.
BERT is a Transformer~\cite{vaswani2017attention} model that achieves results close to state of the art across a variety of NLP tasks~\cite{rogers2020primer}.
We follow the original implementation of BERT using a bidirectional Transformer to build embeddings of each token and then average them into a single vector to represent the code snippet (see the original paper on BERT~\cite{devlin2019bert} for more details).

\subsubsection{AST-based Representation of Code, code2seq}

Compared to texts in natural languages, code has a richer internal structure. This structure can be represented with an abstract syntax tree (AST).
To get an AST, code should be processed by a parser that depends on the programming language and its grammar.
Solid green edges in \Cref{fig:graph} represent AST edges for the code snippet from \Cref{fig:snippet}.

We use code2seq~\cite{alon2018codeseq} as a model that works with AST representation of code.
Code2seq shows state-of-the-art results for code summarization among the models that use information solely from the AST.
In order to represent code, code2seq first samples triples of tokens in two AST leaves and a path between them and encodes the path with an LSTM model~\cite{lstm}. Then the model aggregates the resulting triples into a single vector using the attention mechanism~\cite{luong2015effective}.

\subsubsection{Graph-based Representation of Code, DeeperGCN}

While an AST represents the internal structure of code, it does not capture all of the dependencies between code entities. In order to use the information about them, we can build other representations of code: control flow graphs (CFGs) or program dependence graphs (PDGs).

Control flow graphs describe all paths that can be traversed through the program during its execution. 
Nodes in a CFG are operators and conditions connected by directed edges to indicate the direction of control transfer.
For example, edges between subsequent statements or edges from ``if'' statement to its branches.
Program dependence graphs consist of two types of edges: data dependence edges and control dependence edges. Data edges show dependencies between usages of a variable, \eg they link variable declaration with all other usages. Control edges show that certain code fragments execute depending on condition statements.
For example, $C_{\text{true}}$ connects ``if'' statement with all statements in its ``true'' branch.


All these representations along with the AST may be combined into a single graph.
Figure~\ref{fig:graph} shows an example of such graph representation for code snippet from \Cref{fig:snippet}.
In enriches the existing AST with CFG (shown in dotted red) and PDG (shown in dashed blue) edges.
According to recent works, using such complex graph representations leads to good results in vulnerabilities~\cite{zhou2019devign} or variable misuse~\cite{hellendoorn2019global} detection tasks and code summarization~\cite{fernandes2021structured}.


In order to work with graph-based representation of code, we use the DeeperGCN~\cite{li2020deepergcn} model.
The model consists of multiple layers. First, DeeperGCN embeds graph nodes into numeric vectors $h_v$, then each layer updates the node representation based on the information from the adjacent nodes $N_v$ using the following formula:
\begin{gather}
    h_{v}^k = \sigmoid \left( 
    \sum_{u \in \mathcal{N}_v \cup \{v\}} \frac{1}{c_{u,v}} W^k h_u^{k-1}
    \right),
\end{gather}
where $\sigmoid$ is sigmoid function, $W^k$ is a matrix with learnable weights, and $c_{u, v}$ is a coefficient depending on the degrees of nodes $v$ and $u$.
The number of layers is a hyperparameter of the model that should be fixed in advance. Following the ideas from the CV domain and ResNet architecture in particular~\cite{he2016deep}, DeeperGCN utilizes residual connections between layers in order to improve the quality and speed up the convergence.
\smallskip

As contrastive learning approaches do not impose restrictions on the encoder models, we can use different encoders when applying CL algorithms to source code. In this work, we evaluate the influence of the encoder choice on the results of different CL algorithms in the code clone detection task. 

\section{Experimental setup}\label{sec:setup}

\begin{table*}[!t]
    \centering
    \scalebox{0.97}{
    \begin{tabular}{c|c|c|c|c|c|c}
         \multirow{2}{*}{Dataset} & \multicolumn{2}{c|}{Train} & \multicolumn{2}{c|}{Test} & \multicolumn{2}{c}{Val} \\
    
        & Samples & Classes & Samples & Classes & Samples & Classes \\
    \midrule
        \textsc{POJ-104} & 31,000 & 65 & 10,000 & 23 & 11,000 & 25 \\
        \textsc{Codeforces} & 227,833 & 55,473 & 88,128 & 21,070 & 56,482 & 13,493 \\
    \end{tabular}
    }
    \caption{Statistics for the \textsc{POJ-104} and \textsc{Codeforces} datasets.}
    \label{tab:stats}
\end{table*}

This section defines the experiments we conduct to evaluate the described code representation models paired with different CL approaches. Our goal is to compare the applicability of the trained models to the task of detecting \textsl{Type~IV} clones. We evaluate the models in two settings: detection of functionally equivalent programs on the 
\textsc{POJ-104}~\cite{mou2015convolutional} dataset, and the plagiarism detection task on the dataset of solutions to competitive programming contests held on the \textsc{Codeforces}\footnote{https://codeforces.com} platform. In both tasks, the datasets contain pairs of programs, labeled whether they are clones or not.

\subsection{Clone Detection} 

\textbf{Task description}. In the clone detection task, the model is trained to predict whether two snippets of code are functionally identical. Evaluating the model in this task we gain insights into how the model can comprehend the programs' underlying functionality. 

\textbf{Dataset}. We use one of the most popular datasets in the clone detection field, \textsc{POJ-104}~\cite{mou2015convolutional}, which was previously  used in many clone detection studies~\cite{bui2020infercode, ye2021misim}. \textsc{POJ-104} consists of 104 different algorithms (\eg sorting or string matching algorithms) written in C by the users of the LeetCode\footnote{\url{https://leetcode.com}} platform. There are 500 files per each of the 104 algorithms, which results in a total dataset size of 52,000 files. We consider all the implementations of the same algorithm to be clones. We split the dataset into training, validation, and testing sets by classes (\ie implemented algorithms do not overlap between training/validation/testing sets) in the proportion of $60:20:20$. \Cref{tab:stats} summarizes the dataset's statistics.

\textbf{Metrics}. A common way to measure models' performance in clone detection task is to use F-score~\cite{astnn, wang2018ccaligner}. However, this metric has its problems: it does not consider any ordering of the result, and in a real-world setting it requires tuning the threshold, which is highly sensitive to each particular dataset~\cite{ragkhitwetsagul2018comparison}. Due to this fact, instead of tuning thresholds for F-score, we considered F-score@$R$ that calculates F-score but using only $R$ most relevant samples for each given query. Although such metric handles problem with threshold selection, F-score@$R$ still does not take into account any ordering.
To address this problem we also considered Mean Average Precision at $R$ (MAP@$R$). MAP@$R$ was initially introduced for recommendation tasks~\cite{musgrave2020metric} and is now commonly employed in the clone detection setting~\cite{ye2021misim}. MAP@$R$ measures how accurately a model can retrieve a relevant object from a dataset given a query. It is defined as the mean of average precision scores, each evaluated for retrieving $R$ samples most similar to the query. In our setting, the query set contains all programs from the testing set. We set $R$ to be equal to the number of other programs implementing the same algorithm as a given query program. That is, in \textsc{POJ-104} $R$ equals 500 since each file in this dataset has 500 clones, including the file itself.

\subsection{Plagiarism Detection} 
\textbf{Task description}. The main objective of this task is to identify whether two snippets of code are the plagiarized copies of each other. In other words, the objective is to train a model to be invariant to the transformations of source code, introduced by plagiarism. A decent solution to this task could be applied in the educational settings to fight cheating. 

To obtain a dataset for this task, we designed and implemented a tool \gorshochek which augments the existing dataset with code files that mimic plagiarism.
This choice was made due to the origin of the problem itself: it is very hard to create a manually labeled dataset for plagiarism detection.
When manually collecting such a dataset, positive labels would only correspond to examples when an assessor detected cheating, and therefore, all successful plagiarism attempts would be marked as original code.
To deal with it, we use synthetically generated plagiarism examples. This way, for all pairs of code samples, we know whether they should be classified as clones or not.


\textbf{\gorshochek}.
There were several previous attempts to augment source code by introducing transformations~\cite{devoremcdonald2020mossad, cheers2020detecting, jain2020contrastive, quiring2019misleading}.
Tools that introduce code transformations heavily rely on the syntax of the particular programming language, and  
therefore, cannot be reused with other languages. For example, the tool developed by Jain et al.~\cite{jain2020contrastive} implements transformations only for JavaScript.

Quiring et al.~\cite{quiring2019misleading} created a tool for code transformation that targeted C and C++ languages, which is suitable for our needs. Despite our best efforts, we were unable to reuse it due to the multiple issues we faced when building the project.
However, the set of transformations performed by this tool is very useful.
For these reasons, we decided to develop an easy-to-use and extensible tool of our own. 


In \gorshochek~\cite{gorshochek}, we implement nine different code transformations, they are presented in~\Cref{tab:transformations}. In addition to the transformations already proposed in literature, which we filtered to be 
suitable for the plagiarism detection task, we also added new transformations, which were suggested to us by experts in competitive programming and teaching, who have extensive experience in reading students' code and encountered the most popular patterns for cheating.

\begin{table}[h]
    \centering
    \begin{tabular}{l}
         Transformations \\
    \midrule
        1. Add, remove comments \\
        2. Rename variables \\
        3. Rename functions \\
        4. Swap \texttt{if} and \texttt{else} blocks and change \\ 
        \ \ \ the corresponding \texttt{if} condition \\
        5. Rearrange function declarations \\
        6. Replace \texttt{for} with \texttt{while} \\
        7. Replace \texttt{while} with \texttt{for} \\
        8. Replace \texttt{printf} with \texttt{std::cout} \\ 
        9. Expand macros \\
    \end{tabular}
    \caption{List of implemented transformations in \gorshochek.}
    \label{tab:transformations}
\end{table}

We implemented \gorshochek in C++ using Clang/LLVM's Tooling library (libTooling) --- a set of libraries for traversing, analyzing and modifying ASTs of programs written in C/C++.
While developing the tool, we aimed to make \gorshochek easily extensible with new transformations.
As a result, to add a new transformation one needs to implement just a few interfaces without diving deep into the parser's API.
The tool uses multiprocessing to achieve better performance while working with a large number of code snippets.
To make the tool easily reusable by other researchers and practitioners, we provide a Docker container that already contains all necessary dependencies.

In order to apply \gorshochek for their task, the user should define a configuration through a YAML file.
The configuration includes the number of augmented files for each source file, the list of transformations, and the probability of applying them to code.
Thus, for each file, \gorshochek applies the listed transformations with a certain probability. \Cref{fig:augmentatin} shows an example of augmented code.
In this work, we apply all nine transformations with an equal probability of 0.3, which leads us to an average of three transformations per file.
For each file, we create four augmented versions of it.
The number of augmented files is a trade-off between the dataset's diversity and its size.
Thus, the dataset contains five copies of each file -- the original one and four augmentations -- which are considered pairwise clones.



\begin{figure}
    \centering
    
    \begin{subfigure}{0.45\textwidth}
    \begin{lstlisting}
    #include <stdio.h>
    
    void foobar(int keq) {
    	int lec = 0;
    	int fur = 0;
    	// looooop
    	while (fur < keq) {
    		lec += fur + 1;
    		for(int i =0; i<3; ++i) {
    			fur++;
    			printf("%d", lec);
    		}
    	}
    }
    
    int main(void) {
    	foobar(12);
    	return 0;
    }
    \end{lstlisting}
     \caption{Initial snippet of code}
    \label{lst:source}
    \end{subfigure}

    \begin{subfigure}{0.45\textwidth}
    \begin{lstlisting}
    #include <iomanip>
    #include <iostream>
    #include <stdio.h>
    
    void ai(int ddk) {
      int j = 0;
      int sdd = 0;
      for (; sdd < ddk;) /* 'for' */ {
        j += sdd + 1;
        int tj = 0;
        for (; tj < 3;) /* 'for' */ {
          sdd++;
          std::cout << j;
          ++tj;
        }
      }
    }
    
    int main() {
      ai(12);
      return 0;
    }
    \end{lstlisting}
    \caption{
Snippet of code after augmentation with \gorshochek.
All transformations except swapping \texttt{if-else}, rearranging function declarations, and expanding macros are applied.
}
    \label{lst:aug}
    \end{subfigure}
    \caption{Example of applying \gorshochek to source code}
    \label{fig:augmentatin}
\end{figure}

\textbf{Dataset}. In the scope of this research, we employ a dataset of solutions to competitive programming problems hosted on \textsc{Codeforces}, a platform for coding competitions. All the solutions in this dataset are written in C/C++. We augment the dataset by using \gorshochek.
The final dataset consists on average of 4 plagiarised snippets per each solution from the original \textsc{Codeforces} dataset. We split the dataset into training, validation, and testing sets in the proportion of $60:20:20$. \Cref{tab:stats} summarizes the datasets we use in this work.

\textbf{Metrics}. Similar to the clone detection setting, we evaluate the trained models using F1@$R$ and MAP@$R$. In the plagiarism detection task with \textsc{Codeforces} we chose $R$ to be equal to 5 since each file from the original \textsc{Codeforces} dataset on average has 4 plagiarised copies in the final dataset.

\textbf{Negative samples in CL}. SimCLR and Moco, unlike SwAV, require negative examples (pairs of varying code snippets) along with the positive ones (pairs of clones). The original implementations of SimCLR and Moco assume the elements of a training batch to be augmented versions of different origins.  However, this assumption may be violated in the plagiarism detection task since the model may receive a batch containing multiple plagiarised copies of the same original code snippet. To deal with this problem, a common approach is to embed the information about intra-batch clones into the loss function. Generalized versions of SimCLR and Moco, which take this aspect into account, are called SupCon~\cite{khosla2021supervised} and UniMoco~\cite{dai2021unimoco}, respectively. We use them in our experiments with SimCLR and Moco.

\subsection{Baselines} We compare our models with several baselines, including a non-ML approach Simian,\footnote{https://www.harukizaemon.com/simian/} CCAligner~\cite{wang2018ccaligner}, JPlag~\cite{prechelt2002finding}, and modern ML-based approaches Infercode~\cite{bui2020infercode} and Trans-Coder~\cite{lachaux2020unsupervised}. 

Simian is a commonly used baseline in the code clone detection task, which can handle various programming languages, and showed good results when dealing with \textsl{Type~I-III} clones~\cite{article_Svajlenko}. Simian is a token-based tool, which finds clones ignoring less relevant information (\eg whitespaces, comments, imports). For Simian, we use default hyperparameters and measure the similarity of two programs as a number of lines marked similar by Simian. 

JPlag~\cite{prechelt2002finding} is another popular code clone detection tool. JPlag takes a set of programs, converts each program into a string of canonical tokens (\eg \textsc{BEGIN\_WHILE}, \textsc{END\_METHOD}) and compares these representations pair-wise, computing a total similarity value for each pair. Due to token unification, in practice, JPlag demonstrates robust performance in detecting plagiarism among student program submissions~\cite{misc2016comparison}. In our experiments, we used the tool with its default parameters.

Another non-ML approach that we considered is CCAligner~\cite{wang2018ccaligner}. It utilizes edit distance to calculate a similarity score for code fragments.
While performing generally good with \textsl{Type~I-III} code clones, CCAligner dominates in detecting clones that strongly differ in size.
However, CCAligner is only capable of working with C and Java, which prevents this tool from being used on C++ dataset generated with \gorshochek for the Plagiarism Detection task. In our experiments, we used the tool with its default parameters.

InferCode~\cite{bui2020infercode} is an ML-based approach, based on a novel pre-training method that does not require any labeled data.
In InferCode, the authors utilize code's AST and train the model to find sub-trees of the given tree. This model has shown impressive results in various SE tasks, including clone detection and method name prediction. We extracted embeddings from the final layer of the InferCode network. We use cosine similarity of the embeddings as a measure of similarity of two programs.

The final architecture which we use as a baseline is Trans-Coder~\cite{lachaux2020unsupervised}. Trans-Coder is a Transformer-based model initially designed for the code translation task. Trans-Coder represents the class of models that produce meaningful embeddings of source code because they were pre-trained on a large dataset. The pre-training task of Trans-Coder is code translation, which requires the model to understand the underlying functionality of the program. In this regard, we expect Trans-Coder to show high results in the clone search task as well. As with the InferCode, we extract embeddings from Trans-Coder from its final layer and measure the similarity between code snippets as cosine similarity of their embeddings.

\section{Implementation Details}\label{sec:implementation}

\subsection{Contrastive framework}

\newdimen\figrasterwd
\figrasterwd\textwidth

\begin{figure*}[ht]
\begin{center}
    \includegraphics[width=0.8\hsize]{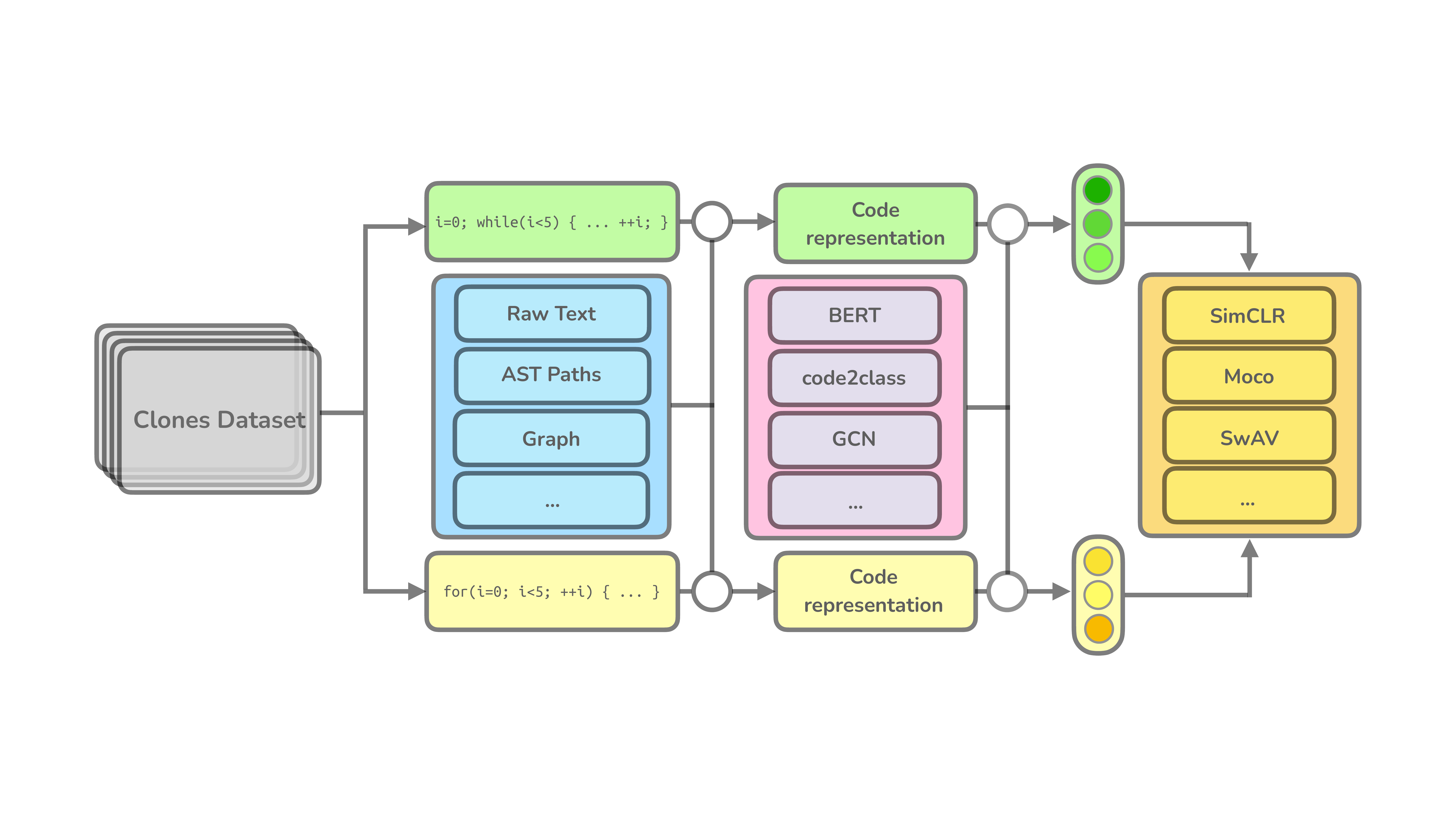}
\end{center}
    \caption{High-level view of the proposed framework. At the first step, we sample a pair of snippets from a dataset. Then, snippets transform into one of the source code representations. Next, we compute embeddings of the respective snippets with the selected model, and finally, feed the embeddings to a contrastive learning algorithm. The framework can be extended by introducing new clone detection datasets, code representation methods, encoder models, and CL approaches.}
    \label{fig:framework}
\end{figure*}

We propose a unified, modular, and extensible open-source framework~\cite{clf} to train and evaluate models in clone detection tasks. Our framework is implemented with PyTorch~\cite{paszke2019pytorch}.
\Cref{fig:framework} presents an overview of the framework. It consists of three core components: (1) building representations of source code, (2) encoding the representation via trainable encoder networks, and (3) training models with a CL approach. 

We make an effort to design the framework to be extensible and reusable. It allows users to experiment with their own code representations, encoder architectures, CL approaches, and datasets for the clone detection task. In order to change any step of the pipeline, the user needs to implement a single Python interface. 
Moreover, we provide an interface to integrate embeddings from pre-trained code comprehension models, allowing researchers and practitioners to benchmark the performance of such embeddings in the clone detection task. The framework already contains all the approaches we describe in this work, which can make it a strong basis for future research on clone detection.


\subsection{Model configurations}

In this subsection we discuss the choice of models' parameters and tools that we use for code processing.

Hyperparameters in machine learning are non-trainable parameters that are usually fixed before training the model. 
Since it is impossible to optimize these parameters during training, their values should either be pre-selected manually or be tuned based on the validation set. In our experiments, we have three sets of hyperparameters which we have to tune: hyperparameters of the CL approaches, of the encoder models, and general training parameters. 

\subsubsection{CL hyperparameters}

\Cref{tab:cl_hyperparams} shows core hyperparameters of the CL approaches. We aim to use default hyperparameters from the original implementations of the respective CL algorithms, but alter them in order to ensure that each model fits on a single Nvidia T4 GPU. For Moco, we reduce the number of negative samples $M$ to 15,360. For SwAV, we choose the number of prototypes $L$ to be 100 for \textsc{POJ-104} since in this task we have only 104 different classes of algorithms, and 1,000 for \textsc{Codeforces}.

\subsubsection{Encoder configurations}

To extract representations of source code, we employ widely adopted tools, which were previously used in ML and SE domains. We select hyperparameters of encoders in such a way that the models have approximately equal capacities (\ie numbers of parameters), and can be trained on a single Nvidia T4 GPU. Additionally, we choose the size of embeddings to be 128 for all models. Next, we describe the details of preprocessing data for each encoder along with their configurations.

\textbf{Text-based representation, BERT}. We tokenize the source code using the publicly available tool YouTokenToMe.\footnote{ \url{https://github.com/VKCOM/YouTokenToMe}} Dictionary generation is done using Byte Pair Encoding~\cite{sennrich-etal-2016-neural}. We crop the input sequence to the maximum length of 386 tokens since BERT has a quadratic memory footprint with respect to the input sequence length. The BERT encoder has 8 heads and 4 layers with the dimension of the feed-forward part equal to 1,024.
    
\textbf{AST-based  representation  of  code,  code2seq}. We obtain path-based representation of code using the  ASTMiner\footnote{ \url{https://github.com/JetBrains-Research/astminer}}~\cite{kovalenko2019pathminer} tool. The encoder we use is essentially a code2seq path encoder, equipped with a 2-layer perceptron classifier for path aggregation. We refer to this encoder as code2class. For the path encoder, we use default parameters from the original implementation~\cite{alon2018codeseq}.
    
\textbf{Graph-based representation of code, DeeperGCN}. We build AST, CFG, and PDG from code snippets using  Joern\footnote{\url{https://joern.io}}~\cite{rodriguez2010graph}, an open-source tool for C/C++ source code analysis. The encoder is a 6-layer DeeperGCN network with default hyperparameters.

\begin{table*}[!htb]
\centering
    \begin{minipage}{.49\linewidth}
    \centering
    \begin{tabular}{c|l|c|c}
         & \multicolumn{1}{c|}{Hyperparameters} & POJ-104 & Codeforces  \\
    \midrule
        \multicolumn{1}{c|}{SimCLR} &  \multicolumn{1}{l|}{Softmax temperature, $\tau$} & $0.1$ & $0.1$ \\
    \midrule
        \multicolumn{1}{c|}{\multirow{3}{*}{Moco}} & \multicolumn{1}{l|}{Encoder momentum, $m$} & $0.999$  & $0.999$ \\
        \multicolumn{1}{l|}{} & \multicolumn{1}{l|}{Buffer size, $M$} & $15\,360$ & $15\,360$ \\
        \multicolumn{1}{l|}{} & \multicolumn{1}{l|}{Softmax temperature, $\tau$} & $0.07$ & $0.07$ \\
    \midrule
        \multicolumn{1}{c|}{\multirow{2}{*}{ SwAV}} & \multicolumn{1}{l|}{Number of prototypes, $L$} & $100$  & $1000$ \\
        \multicolumn{1}{l|}{} & \multicolumn{1}{l|}{Softmax temperature, $\tau$} & $0.1$ & $0.1$ \\
    \end{tabular}
    \caption{Hyperparametrs of the CL aproaches.}
    \label{tab:cl_hyperparams}
    \end{minipage}
    \begin{minipage}{.49\linewidth}
    \centering
    \begin{tabular}{c|l|c|c}
         & \multicolumn{1}{c|}{Hyperparameters} & POJ-104 & Codeforces  \\
    \midrule
        \multicolumn{1}{c|}{\multirow{5}{*}{BERT}} & \multicolumn{1}{l|}{Vocabulary size, $V$} & $20\,000$  & $40\,000$ \\
        \multicolumn{1}{l|}{} & \multicolumn{1}{l|}{Max sequence length, $L_{\text{max}}$} & $386$ & $386$ \\
        \multicolumn{1}{l|}{} & \multicolumn{1}{l|}{Number of heads, $N_{\text{heads}}$} & $8$ & $8$ \\
        \multicolumn{1}{l|}{} & \multicolumn{1}{l|}{Number of layers, $N_{\text{layers}}$} & $4$ & $4$ \\
        \multicolumn{1}{l|}{} & \multicolumn{1}{l|}{Feadforward size, $D_{\text{feedforward}}$} & $1024$ & $1024$ \\
    \midrule
        \multicolumn{1}{c|}{\multirow{3}{*}{code2class}} & \multicolumn{1}{l|}{Max context, $N_{\text{context}}$} & $200$  & $200$ \\
        \multicolumn{1}{l|}{} & \multicolumn{1}{l|}{Path length, $L_{\text{path}}$} & $9$ & $9$ \\
        \multicolumn{1}{l|}{} & \multicolumn{1}{l|}{Classifier layers, $N_{\text{clf layers}}$} & $2$ & $2$ \\
    \midrule
        \multicolumn{1}{c|}{GCN} & \multicolumn{1}{l|}{Number of layers, $N_{\text{layers}}$} & $6$  & $6$ \\
    \end{tabular}
    \caption{Hyperparametrs of the encoder models.}
    \label{tab:encoder_hyperparams}
    \end{minipage}
\end{table*}

\subsubsection{Training details}

In our experiments, we observe that some models are extremely sensitive to the learning rate.
In this regard, we conduct a grid search~\cite{liashchynskyi2019grid} for them in each experiment.
We search for learning rates among the values of $\{10^{-2}, 10^{-3}, 10^{-4}, 10^{-5}\}$. \Cref{tab:params} presents the best values we obtained for all the models.

For all our experiments, we use a fixed batch size $N$ equal to 80.
Since in most CL approaches the number of negative examples is proportional to the batch size, we choose the batch size as big as possible to fit the computational resources we use.
We select the best model using the validation set and then apply it to the testing set in order to compare the models' quality.

\begin{table*}[!ht]
    \centering
    \begin{tabular}{cccc|ccc}
         & SimCLR & SwAV & Moco & SimCLR & SwAV & Moco  \\
    \midrule
         \multicolumn{1}{c|}{BERT} & $10^{-4}$  & $10^{-5}$ & $10^{-3}$ & $10^{-3}$ & $10^{-5}$ & $10^{-3}$ \\
         \multicolumn{1}{c|}{code2class} & $10^{-5}$ & $10^{-5}$ & $10^{-4}$ & $10^{-3}$ & $10^{-4}$ & $10^{-3}$  \\
         \multicolumn{1}{c|}{GCN} & $10^{-2}$ & $10^{-4}$ & $10^{-5}$ & $10^{-3}$ & $10^{-4}$ & $10^{-3}$  \\
    \end{tabular}
    \caption{Optimal learning rate values, chosen using grid search.}
    \label{tab:params}
\end{table*}

\subsection{Plagiarism Detection}

According to prior research, selection of negative examples plays an important role in the convergence of CL approaches~\cite{robinson2021contrastive}.
Indeed, with our first series of experiments on the \textsc{Codeforces} dataset, we observed that the models rapidly overfitted on the training data, \ie showed high scores on the training set but demonstrated much worse performance on the validation set.
We attribute this result to the fact that the models received solutions of different problems as input, which caused them to learn to distinguish the functionality of code fragments instead of solving the plagiarism detection task.
Due to this fact, we enforce batches in the experiments on the \textsc{Codeforces} dataset to consist of solutions to the same problem.
This issue does not apply to the \textsc{POJ-104} dataset, as in this case we consider all solutions to the same problem to be clones. 

\section{Evaluation Results}\label{sec:results}

\begin{table*}[ht]
    \centering
    \scalebox{0.97}{
    \begin{tabular}{cccc|ccc||ccc|ccc}
         \multicolumn{1}{c|}{Model} & \multicolumn{3}{c|}{POJ-104, MAP@500} &  \multicolumn{3}{c||}{POJ-104, F1@500} & 
         \multicolumn{3}{c|}{Codeforces, MAP@5} &
         \multicolumn{3}{c}{Codeforces, F1@5} \\
    \midrule
        \multicolumn{1}{c|}{CCAligner} & \multicolumn{3}{c|}{$3.35$} &  \multicolumn{3}{c||}{$7.81$} &
        \multicolumn{3}{c|}{--} &
        \multicolumn{3}{c}{--} \\
        \multicolumn{1}{c|}{Simian} & \multicolumn{3}{c|}{$0.6$} &  \multicolumn{3}{c||}{$3.16$} &
        \multicolumn{3}{c|}{$32.59$} &
        \multicolumn{3}{c}{$40.69$} \\
        \multicolumn{1}{c|}{JPlag} & \multicolumn{3}{c|}{$12.09$} &  \multicolumn{3}{c||}{$18.82$} &
        \multicolumn{3}{c|}{$38.88$} &
        \multicolumn{3}{c}{$46.58$} \\
        \multicolumn{1}{c|}{InferCode} & \multicolumn{3}{c|}{$16.25$} &  \multicolumn{3}{c||}{$27.87$} &
        \multicolumn{3}{c|}{$43.01$} &
        \multicolumn{3}{c}{$44.50$} \\
        \multicolumn{1}{c|}{Trans-Coder C++ to Python} & \multicolumn{3}{c|}{$48.39$} &  \multicolumn{3}{c||}{$55.54$} &
        \multicolumn{3}{c|}{$42.43$} &
        \multicolumn{3}{c}{$43.64$}\\
        \multicolumn{1}{c|}{Trans-Coder C++ to Java} & \multicolumn{3}{c|}{$49.39$} &  \multicolumn{3}{c||}{$56.41$} &
        \multicolumn{3}{c|}{$42.68$} &
        \multicolumn{3}{c}{$43.91$}\\
    \\
         & \multicolumn{12}{c}{Contrastive}\\
    \cmidrule(l){2-13}
         & SimCLR & SwAV & Moco & SimCLR & SwAV & Moco & SimCLR & SwAV & Moco & SimCLR & SwAV & Moco \\
    \midrule
         \multicolumn{1}{c|}{BERT} & $35.38$  & $0.44$ & $32.24$ & $48.01$	& $5.03$ & $44.09$ & $1.21$ & $0.38$ & $44.66$ & $1.58$ & $0.43$ & $48.94$ \\
         \multicolumn{1}{c|}{code2class} & $36.59$ & $37.14$ & $33.09$ & $47.32$ & $48.19$ & $45.51$ &
         $40.70$ & $31.23$ & $34.11$ & $42.65$ & $33.28$ & $36.15$  \\
         \multicolumn{1}{c|}{GCN} &
         $\mathbf{55.75}$ &	$\underline{55.16}$ & $52.17$ & $\mathbf{62.84}$	& $\underline{62.59}$ & $60.32$ & $\underline{58.94}$ & $56.19$ & $\mathbf{58.99}$ & $\underline{59.23}$	& $57.19$	& $\mathbf{59.27}$\\
    \end{tabular}
    }
    \caption{Experiment results for all the studied models on two datasets: \textsc{POJ-104} and \textsc{Codeforces}. The values are MAP@$R$ and F1@$R$, with $R$ being 500 for the POJ-104 dataset and 5 for the \textsc{Codeforces} one.}
    \label{tab:results}
\end{table*}

We trained all the models and performed the hyperparameter search on a single Nvidia Tesla T4 GPU.
\Cref{tab:results} presents the results of the models' evaluation on the testing parts of both \textsc{POJ-104} and \textsc{Codeforces} datasets.
In both experiments, all CL algorithms showed the best results when used with the DeeperGCN model. It suggests that information contained in the graph representation of code turns out to be very useful when solving the clone detection problem.

The BERT and code2class models demonstrate worse performance with a significant margin. Notably, the BERT model did not converge at all in three out of six experiments, even though we performed a grid search to identify the suitable learning rate.

\subsection{BERT Convergence}

Three out of six experiments we performed with the BERT model did not converge at all. It means that the optimization procedure did not find model parameters that would allow the model to extract embeddings which are similar for the semantically equivalent code snippets. For Moco, even though we got adequate results for both datasets, only specific learning rate values led to the optimization convergence. For SwAV on both datasets and for SimCLR on \textsc{Codeforces}, all the learning rate values led to convergence failure. Based on the experiments that produced more adequate results, we draw the conclusion that the usage of BERT in a contrastive learning framework is very sensitive to the choice of hyperparameters.

Another conclusion that can be drawn from the experiments conducted with the BERT encoder and Moco is that Moco is way more robust than other CL algorithms. According to Liu et al.~\cite{liu2019roberta}, using large batch sizes during training usually positively affects the BERT model.
Since Moco utilizes a queue of previously processed code snippets to treat them as negative examples, the number of samples processed by the model at each training step is several times higher than that for SimCLR. 

\subsection{Clone Detection Task}

In the clone detection task with the \textsc{POJ-104} dataset, we obtain the best results when using DeeperGCN paired with SimCLR and SwAV, while code2class and BERT showed nearly identical (and substantially lower) scores. It can be attributed to the fact that DeeperGCN works with a richer representation of code, which by design contains more information. Notably, as we want to analyze the underlying semantics of the program, graph-based models are more resistant to changes in code, as long as these changes do not break the code semantics: \eg renaming, reordering of operations.

Out of the three studied CL algorithms, SimCLR performed best with BERT and GCN or on par with other CL methods with code2class. The superior performance of SimCLR and SwAV can be attributed to the fact that these CL approaches, in contrast to Moco, encode both objects with a shared network. In this regard, the encoder model receives updates from each element of the batch. On the other hand, such training scheme has proven to be less stable in experiments with BERT, which has a larger number of parameters.

As expected, Simian and CCAligner failed to achieve good quality when detecting \textsl{Type-IV} clones. Both approaches measure similarity between files based on their tokens, and for different programs implementing the same algorithm in the \textsc{POJ-104} dataset, the overlap of tokens (even computed with some heuristics) can be extremely small. However, another classical approach, JPlag, surpassed others, indicating that token unification is the most accurate approach in detecting \textsl{Type-IV} clones among the token-based tools we compared.

Interestingly, the embeddings learned by Trans-Coder in code-to-code translation tasks turned out to be nearly as useful in clone detection tasks as the ones learned in the CL setting.
It shows that pre-training with the task of code translation indeed produces models that can successfully extract the underlying functionality of source code.

In contrast, InferCode showed unexpectedly low results, which can be attributed to the fact that the pre-training objective of InferCode made it robust only to minor transformations of code. However, it is still challenging for the model to find similarities among the significantly different programs from \textsc{POJ-104}.

\subsection{Plagiarism Detection Task}

The \textsc{Codeforces} dataset that we used for the plagiarism detection task considers code snippets to be clones when one of them was made from another with a number of pre-defined transformations (see~\Cref{tab:transformations}). Thus, in order to be considered clones in this dataset, pairs of code snippets should not only be functionally equivalent (as regular \textsl{Type-IV} clones) but also share more code-level similarities. This difference impacts the results significantly compared to the \textsc{POJ-104} dataset.

Similar to the previous experiment with \textsc{POJ-104}, DeeperGCN performs best with all the CL algorithms. When detecting plagiarism, the gap in quality between graph-based models and other approaches becomes even more significant compared to the clone detection case. As with clone detection, graph representation turns out to be the most robust to variations in the implementation, as it pays more attention to the internal structure of the solution, which is harder to change.

BERT successfully converged only in combination with the Moco approach. However, in this case, it achieved decent results, outperforming all other approaches except for graph-based models. Poor performance of the BERT encoder coupled with SimCLR and SwAV again supports the hypothesis that the momentum encoding scheme is more robust in terms of huge models. 

In contrast to the clone detection case, the approaches based on the code2class encoder perform significantly worse than the others. In order to represent a code fragment, code2class samples a fixed number of paths from the AST. When the code fragment is large (and in this experiment, the model takes a whole program as input), a particular choice of sampled paths strongly influences the prediction. Even for  the nearly identical files, if the sampled sets of paths significantly differ, the model might fail to identify the similarity. A possible way to mitigate this issue is the development of better techniques for choosing which paths to sample from the syntax tree.

As the plagiarism detection task required models to identify transformed files, Simian, JPlag, and InferCode significantly improved their performance compared to the code clone detection. We attribute this to the fact that in this setting code snippets considered to be clones share more textual similarities than in the previous setting.

In contrast to other approaches, the performance of Trans-Coder models dropped compared to the \textsc{POJ-104} dataset. Since we used embeddings produced by these models without any fine-tuning for the specific task, these models continued to consider files that they believed shared the same functionality to be clones. However, for the plagiarism detection task, the models had to take into account the implementation-level similarity, as we considered only the transformed snippets to be clones.
\smallskip

Overall, as expected, ML-based models significantly outperformed token-based Simian, CCAligner, and JPlag in both tasks. The pre-trained Trans-Coder have indeed learned meaningful features from the code translation task and achieved results comparable to the models trained in the CL paradigm. In both clone and plagiarism detection tasks, graph-based representations led to the best performance. In terms of contrastive learning approaches, we conclude that SimCLR and SwAV performed on par or even better than Moco. This result can be attributed to the fact that the encoder in SimCLR and SwAV receives more information during training than the encoder in Moco due to the fact that the latter one uses only a single encoder. However, Moco shows to be more robust combined with BERT and even outperforms the pre-trained Trans-Coder in the plagiarism detection task.


\section{Related work}\label{sec:related-work}

Searching for code clones is an important yet not fully solved task in the Software Engineering domain.
According to the survey by Ain et al.~\cite{ain2019systematic}, there are plenty of different approaches to detecting code clones that differ in the way the code is processed, in target languages, or supported platforms.
While approaches targeting \textsl{Type~I-III} clones show decent performance, the detection of \textsl{Type~IV} clones turns out to be still very difficult for most of the models.

The existing works on the detection of \textsl{Type~IV} clones commonly used machine learning techniques~\cite{ye2021misim, wang2020detecting, ling2020hierarchical, li2019graph, bui2020infercode, jain2020contrastive}. These works differ in two major ways: by the ML approaches they use and by the way they extract information from source code. 

From the perspective of ML approaches, in our work we mainly focus on the contrastive learning paradigm for training the models. In recent years, contrastive learning proved to be useful in detecting similar objects in both computer vision and natural language processing~\cite{chen2020mocov2, chen2020simple, chen2021empirical, giorgi2021declutr, rethmeier2021primer, xie2021selfsupervised}. Driven by this success, CL approaches already found application in the task of clone detection in program code~\cite{jain2020contrastive,funcgnn,wang2020detecting,ye2021misim}.

The study by Jain et al.~\cite{jain2020contrastive} focused on different types of model pre-training on source code, while studying a single contrastive learning algorithm.
Ye et al. presented another CL approach to clone detection~\cite{ye2021misim}. The authors introduced CASS, a new graph representation of code, and showed that enriching AST with additional edges leads to a significant performance improvement in the clone detection task.
However, the CL algorithm utilized in this work was previously outperformed by several modern approaches~\cite{jaiswal2021survey}.
A recent comparative study of code representations assessed eight ML models, including text-based and AST-based ones on several tasks, including code clone detection~\cite{han2021comparison}. However, the authors trained the models to solve a classification task rather than using a contrastive learning paradigm.

In contrast to the described studies, we compare multiple modern contrastive learning algorithms between themselves, as well as with solutions that do not involve contrastive learning: Simian, InferCode~\cite{bui2020infercode}, and Trans-Coder~\cite{lachaux2020unsupervised}.

From the perspective of code representations, we explore which representations work better for encoders in CL methods. In previous works on code clone detection, researchers represented code as a sequence of tokens~\cite{jain2020contrastive, feng2020codebert}, as Abstract Syntax Trees and its derivatives~\cite{bui2020infercode, alon2018codeseq}, and as more complex graph structures such as CFG and PDG~\cite{ye2021misim, bennun2018neural, 10.1145/3395363.3397362}. Following these works, as encoders we selected three models that represent code differently: BERT~\cite{devlin2019bert}, code2seq~\cite{alon2018codeseq}, and DeeperGCN~\cite{li2020deepergcn}.

\section{Conclusion}\label{sec:conclusion}

With this work, we present a comprehensive study of modern contrastive learning approaches in combination with different source code representations and encoder models in the setting of the code clone detection task. We demonstrate that graph-based models indeed show better results than text-based models. Moreover, we show that the SimCLR and SwAV contrastive approaches can achieve higher scores, while Moco is generally more robust. Finally, we demonstrate that for the code clone detection task, the embeddings learned within various pre-training objectives (\eg AST subtree prediction or code translation) can serve as a strong baseline, even compared with the models trained in a supervised manner.

We introduce a novel tool called \gorshochek for augmentation of source code. In \gorshochek, we implemented nine source code transformations, simulating plagiarism. We designed the tool to be extensible and easy to use. In addition, we propose a highly extensible framework for training and evaluation of ML models for the clone detection task.  We believe that the tool and the framework developed in the scope of this work can serve as a basis for future development in the clone detection field. To this end, we make all the code, tools, datasets, and trained models publicly available.

We make both the evaluation framework and \gorshochek publicly available~\cite{clf,gorshochek}. We will also make all the datasets we used public upon the paper acceptance. We believe that these artifacts will serve as a basis for future development in the clone detection task. Possible future work includes the reuse of the embeddings learned in the contrastive learning paradigm for other tasks, \eg refactoring recommendation or documentation generation, adding new code transformations for more robust plagiarism detection, and extension of our work to other programming languages.

\bibliographystyle{ACM-Reference-Format}
\bibliography{papers}

\end{document}